\documentclass[prb,showpacs,twocolumn,superscriptaddress,floatfix,amsmath]{revtex4}
\usepackage[english]{babel}
\usepackage{graphicx}
\def\ket#1{\left| #1 \right\rangle}             
\begin{document}
\title{A proof of the Kramers degeneracy of transmission eigenvalues from antisymmetry of the
scattering matrix}
\author{J.~H.~Bardarson}
\address{Instituut-Lorentz, Universiteit Leiden, P.O. Box 9506,
2300 RA Leiden, The Netherlands}
\date{May 2008}
\begin{abstract}
In time reversal symmetric systems with half integral spins (or more concretely, systems with an antiunitary symmetry that squares to -1
and commutes with the Hamiltonian) the transmission eigenvalues of the scattering matrix come in pairs.  We present a proof of this fact that is valid both for even and odd number of modes and relies solely on the antisymmetry of the scattering matrix imposed by time reversal symmetry. 
\end{abstract}
\maketitle

\section{Introduction}

In the early days of quantum mechanics, Kramers noticed a highly consequential feature of the Schr{\"o}dinger equation with
spin~\cite{Kra30}. In the
absence of a magnetic field, he found a transformation $T$ such that if $\ket{\psi}$ is an eigenstate with energy $E$, then
$T\ket{\psi}$ is also an eigenstate. Furthermore, if the system has half integral spin, these two eigenstates are orthogonal and the
spectrum is twofold degenerate. This degeneracy is the celebrated {\it Kramers degeneracy} (of energy eigenvalues). 

A couple of years after Kramers' discovery, Wigner pointed out that the transformation $T$ is simply time reversal and that the degeneracy
is a manifestation
of time reversal symmetry~\cite{Wig32}. In general, Wigner has shown that all symmetry operations can be represented by either an unitary
(and linear)
or an antiunitary (and antilinear) operator that commutes with the Hamiltonian~\cite{Wigner}. In the case of time reversal, the operator $T$ is antiunitary.
Furthermore, the square of the time reversal operator equals plus or minus one~\cite{Wigner}. For half-integral spin $T^2 = -1$ while for
integral spin $T^2 = 1$. Thus, the mathematical conditions for Kramers degeneracy are that there exists an antiunitary operator
$T$ with $T^2 = -1$, that commutes with the Hamiltonian $H$ ($T^2 = -1$ is needed to guarantee that $\ket{\psi}$ and $T\ket{\psi}$ are
orthogonal). Time reversal symmetry is just one example of a symmetry that is represented by an antiunitary operator, but
since all antiunitary operators can be written as a product of an unitary operator and the time reversal operator one generally just
speaks of time reversal. 

A more recently appreciated consequence of time reversal symmetry is the fact that in two terminal scattering of systems with half-integral spin, the transmission eigenvalues
(eigenvalues of the transmission matrix product $tt^\dagger$) also come in Kramers degenerate pairs. This is a less intuitive result than
for the energy eigenvalues. In fact, one expects an electron moving in some direction to have the same energy as an electron moving in the opposite
direction. In terms of scattering, however, time reversal connects incoming states to outgoing states, making it less obvious that there
should be any degeneracy in the transmission eigenvalues. 

In this paper we give a new proof of this Kramers degeneracy of the transmission eigenvalues. There are at least two reasons why this might be beneficial. Firstly, the proofs of this statement that exist in the literature~\cite{Mel91, Mac92} are not given directly in terms of the scattering matrix through which the transmission eigenvalues are obtained, but rather by use of the transfer matrix. While this is equivalent, it  tends to obscure and makes the proofs by far less accessible than the proof of the corresponding theorem for energy levels~\cite{Dim06}.  Secondly, these proofs rest on the quaternionic structure of the transfer and scattering matrices and are thus strictly only applicable to systems with even number of modes, while Kramers degeneracy is also present for systems with odd number of modes.

\section{The proof}
The proof consists of two steps. In the first step we introduce a basis in which the scattering matrix is found to be antisymmetric. In the second step we use the resulting antisymmetry of the matrix of reflection amplitudes to show that the transmission eigenvalues necessarily come in pairs.

Consider thus a two terminal scattering setup with $N_{L}$ ($N_R$) propagating modes in the left (right) lead. Assume further that the system
has a time reversal (antiunitary) symmetry $T$ that square to minus one, $T^2=-1$. We denote the incoming modes on
the left (right) with $\ket{n}$ ($\ket{m}$). The outgoing modes on the left (right) are then $T\ket{n}=\ket{Tn}$ ($T\ket{m}=\ket{Tm}$).
Since $\langle n|Tn\rangle = 0$ as a consequence of the properties of $T$, it is straightforward to show (for example using the
Gram-Schmidt procedure) that the basis \{$\ket{n},\ket{Tn}$\} can
always be chosen to be orthogonal. A general scattering state $\ket{\psi}$ will thus have the form
\begin{equation}
  \ket{\psi} = 
    \sum_{n=1}^{N_L} (c_n^{\rm{in,L}}\ket{n} + c_n^{\rm{out,L}}\ket{Tn})
  \label{eq:ScatteringState}
\end{equation}
in the left lead, and similar in the right lead (with $n\rightarrow m$ and $L\rightarrow R$). The vectors of coefficients $c^{\rm in}$ and $c^{\rm out}$ are related by the scattering
matrix $S$,
\begin{equation}
\begin{pmatrix}
  c^{\rm out,L} \\ c^{\rm out,R} 
\end{pmatrix}
  = S
\begin{pmatrix}
  c^{\rm in,L} \\ c^{\rm in,R} 
 \end{pmatrix}  =  
\begin{pmatrix}
      r & t^\prime \\ t & r^\prime
 \end{pmatrix}
\begin{pmatrix}
   c^{\rm in,L} \\ c^{\rm in,R} 
\end{pmatrix}.
  \label{eq:S}
\end{equation}
Due to time reversal symmetry,  
\begin{equation}
  T\ket{\psi} = 
    \sum_{n=1}^{N_L} [(c_n^{\rm{in,L}})^*\ket{Tn} - (c_n^{\rm{out,L}})^*\ket{n}]
  \label{eq:TScatteringState}
\end{equation}
is also a scattering state with the same energy. 
That means that 
\begin{equation}
\begin{pmatrix}
  (c^{\rm in,L})^* \\ (c^{\rm in,R})^* 
\end{pmatrix}
  = S
\begin{pmatrix}
   -(c^{\rm out,L})^* \\ -(c^{\rm out,R})^* 
\end{pmatrix}. 
  \label{eq:ST}
\end{equation}
Using unitarity of $S$ we conclude, by comparison with Eq.~\eqref{eq:S}, that in our chosen basis $S$ is antisymmetric
\begin{equation}
  S^T = -S.
  \label{eq:Santisymmetric}
\end{equation}

We note that if the outgoing states come in pairs, such that if $T\ket{n}$ is an outgoing state $i\sigma_2T\ket{n}$ is also an outgoing
state, it is most common to pair $\ket{n}$ and $i\sigma_2T\ket{n}$ rather than $\ket{n}$ and $T\ket{n}$ as we have done (this way the
pairs have the same spin, while in our case it is opposite). In this case the
scattering matrix is found to be self-dual~\cite{Bee97} $\sigma_2S^T\sigma_2 = S$. It is, however, not always possible to find such pairing
(one example where it is possible is the usual case of metallic leads with electrons of spin half and no spin-orbit coupling), and we will therefore
use the more general representation~\eqref{eq:ScatteringState} in which $S$ is antisymmetric.

As a consequence of the antisymmetry of the scattering matrix, the $N_L \times N_L$ matrix of reflection amplitudes $r$ is also
antisymmetric. Being antisymmetric $r$ can be decomposed~\cite{You61,Sch01} as $r = W^TDW$, with $W$ a unitary matrix and 
\begin{equation}
  D = \Sigma_1 \oplus \Sigma_2 \oplus \cdots \oplus \Sigma_k \oplus 0 \oplus \cdots \oplus 0,
  \label{eq:SVDantisymmetric}
\end{equation}
where $2k = {\rm rank}\ r$, $\oplus$ denotes the direct sum  and 
\begin{equation}
  \Sigma_j =
 \begin{pmatrix}
    0 & \lambda_j \\ -\lambda_j & 0
\end{pmatrix}
  ,\quad \lambda_j > 0, \quad j=1,\cdots, k.
  \label{eq:Sigmar}
\end{equation}
In other words,  $D$ is block diagonal with $k$ $2\times2$ nonzero blocks $\Sigma_j$ and $N_L - 2k$ $1\times 1$ zero blocks $0$. Clearly
if there are odd number of modes (i.e.\ the dimension of $r$ is odd) there is at least one zero block in the
sum~\eqref{eq:SVDantisymmetric}. Using this result, we find that
\begin{equation}
  r^\dagger r = W^\dagger D^TD W. 
  \label{eq:rdaggerr}
\end{equation}
But since 
\begin{equation}
  \Sigma_j^T\Sigma_j = 
\begin{pmatrix}
    \lambda_j^2 & 0 \\ 0 & \lambda_j^2  
\end{pmatrix},
\end{equation}
the matrix $D^TD$ is diagonal. We have thus managed to diagonalize $r^\dagger r$ and found that its eigenvalues come in pairs. Due to unitarity of $S$, $\openone - r^\dagger
r$ and $t^\dagger t$ have the same eigenvalues. The transmission eigenvalues are thus twofold degenerate, plus (if the
number of modes is odd) one transmission eigenvalue equal to unity (perfect transmission~\cite{And98}). This is the Kramers degeneracy of
transmission eigenvalues.

\vspace{0.3cm}
\section{Conclusion}
In conclusion, we have presented, using only the symmetry of the scattering matrix, a concise proof of the fundamental fact that in the presence of time reversal symmetry and for half-integral spin the transmission eigenvalues of the two terminal
scattering matrix come in (Kramers) degenerate pairs.

\section*{ACKNOWLEDGMENTS}
I thank C.\ W.\ J.\ Beenakker for helpful discussions.

\end{document}